\documentclass[12pt,epsf,epsfig,psfig]{article}
\usepackage{graphicx}
\usepackage{epsfig}
\def\J{$J/\psi$}

\def\X{$\chi_c$}
\def\x{\chi}
\def\P{$\psi'$}

\def\U{$\Upsilon$}

\def\e{\epsilon}

\def\m{m_{\rm th}}

\def\be{\begin{equation}}
\def\ee{\end{equation}}

\def\lsim{\raise0.3ex\hbox{$<$\kern-0.75em\raise-1.1ex\hbox{$\sim$}}}
\def\gsim{\raise0.3ex\hbox{$>$\kern-0.75em\raise-1.1ex\hbox{$\sim$}}}
\def\NP{{ Nucl.\ Phys.\ }}
\def\PL{{ Phys.\ Lett.\ }}
\def\PR{{ Phys.\ Rev.\ }}

\def\ZP{{ Z.\ Phys.\ }}

\begin{document}

\hfill BI-TP 2008/06

\vskip 2cm

\centerline{\Large \bf The Thermodynamics of Quarks and Gluons$^*$}

\vskip 1.5cm 

\centerline{\large \bf Helmut Satz} 

\vskip 0.5cm

\centerline{Fakult\"at f\"ur Physik, Universit\"at Bielefeld, Germany}
 
\vskip2.5cm

\centerline{\bf Abstract:}

\bigskip

This is an introduction to the study of strongly interacting matter.
We survey its different possible states and discuss the transition from
hadronic matter to a plasma of deconfined quarks and gluons. Following
this, we summarize the results provided by lattice QCD finite temperature 
and density, and then investigate the nature of the deconfinement 
transition. Finally we give a schematic overview of possible ways to 
study the properties of the quark-gluon plasma. 

\vfill

* Lecture given at the {\sl QGP Winter School}, Jaipur/India, Feb.\ 1, 2008;
to appear in 

\centerline{\sl Springer Lecture Notes in Physics}

\newpage

\section{Prelude} 

The fundamental questions of physics appear on two levels,
the microscopic and macroscopic. We begin by asking:

\begin{itemize}
\vspace*{-0.2cm}
\item{What are the ultimate constituents of matter?}
\vspace*{-0.3cm}
\item{What are the basic forces between these constituents?}
\end{itemize} 

\vspace*{-0.2cm}

Given the basic building blocks and their interactions,
we want to know:

\begin{itemize}
\vspace*{-0.2cm}
\item{What are the possible states of matter?}
\vspace*{-0.3cm}
\item{How do transitions between these states take place?}
\end{itemize} 

\vspace*{-0.2cm}

How far have we advanced today in our understanding of these different 
aspects?

\medskip

According to our present state of knowledge, the ultimate constituents
are quarks, leptons, gluons, photons, intermediate vector bosons ($Z/W^{\pm}$)
and Higgs bosons - in a conservative count (no antiparticles etc.),
sixteen in all, with gravitation not yet in the game. 

\medskip

Their interactions were orignally classified as strong, electromagnetic, 
weak and gravitation, leaving a more general scheme as a challenge.
The first unification brought electroweak theory, the second combined 
this with strong interactions to the standard model. The origin of all 
the different basic constituents, as well as the role of gravitation, 
are still open, waiting for the theory of everything (TOE).

\medskip

In ancient times, the basic states of matter were earth, water, air and fire;
today we have solids, liquids, gases and plasmas. In addition, there now
is a multitude of others: insulators, conductors and superconductors, fluids 
and superfluids, ferromagnets, spin glasses, gelatines and many more. 
And the question of the possible states of matter brings us to a new kind 
of physics; the knowledge of the elementary constituents and their 
interactions in general does not predict the structure of the possible 
complex states of many constituents. 

\medskip

The study of complex systems becomes even more general, less dependent
on the microstructure, when we ask for the transitions between the different 
states. We have phase transitions, depending on the singular behaviour of 
the partition function determined by the respective dynamics, as well as
clustering and percolation transitions, determined by the connectivity 
aspects of the system. But we then find that scaling and renormalization 
concepts lead to a universal description of critical phenomena, and
critical exponents define universality classes which contain quite 
different interaction forms. 
 
\medskip

When we study strongly interacting matter, we are therefore led to
aspects which are relevant not only to QCD, but to the understanding 
of complex systems in general.
 
\section{States of Strongly Interacting Matter}

\begin{figure}[htb]
\centerline{\psfig{file=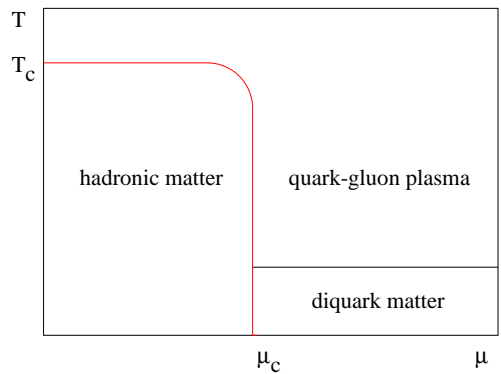,width=6.5cm}}
\caption{The phase diagram of QCD} 
\label{phase}
\end{figure}

What happens to strongly interacting matter in the limit of
high temperatures and densities? This question has fascinated
physicists ever since the discovery of the strong force and the 
multiple hadron production it leads to. Let us look at some of the 
features that have emerged over the years.
\begin{itemize}
\item{Hadrons have an intrinsic size, with a radius $r_h \simeq
1$ fm, and hence a hadron needs a space of volume $V_h \simeq (4\pi
/3)r_h^3$ in order to exist. This suggests a limiting
density $n_c$ of hadronic matter \cite{Pomeranchuk}, with $n_c = 1/V_h
\simeq 1.5~n_0$, where $n_0 \simeq 0.17$ fm$^{-3}$ denotes the 
density of normal nuclear matter.}
\vspace*{-0.2cm}
\item{Hadronic interactions provide abundant resonance production, and 
the resulting number $\rho(m)$ of hadron species increases
exponentially as function of the resonance mass $m$,
$\rho(m) \sim \exp(b~\!m)$. Such a form for $\rho(m)$ appeared first
in the statistical bootstrap model, based on self-similar resonance
formation or decay \cite{Hagedorn}. It was then also obtained in the 
more dynamical dual resonance approach \cite{DRM}. In hadron
thermodynamics, the exponential increase of the resonance degeneracy results 
in an upper limit for the temperature of hadronic matter, 
$T_c = 1/b \simeq 150-200$ MeV \cite{Hagedorn}.}
%\vspace*{-0.6cm}
\item{What happens beyond $T_c$? In QCD, hadrons are dimensionful 
color-neutral bound states of more basic pointlike colored quarks and
gluons. Hadronic matter, consisting of colorless constituents of hadronic
dimensions, can therefore turn at high temperatures and/or densities
into a quark-gluon plasma of pointlike colored quarks and gluons as 
constituents \cite{C-P}. This deconfinement transition leads to a 
colour-conducting state and thus is the QCD counterpart of the 
insulator-conductor transition in atomic matter \cite{HS-Fort}.}
\vspace*{-0.2cm}
\item{A shift in the effective constituent mass is a second transition
phenomenon expected from the behavior of atomic matter. At $T=0$, 
in vacuum, quarks dress themselves with gluons to form the constituent 
quarks that make up hadrons. As a result, the bare quark mass $m_q \sim 0$
is replaced by a constituent quark mass $M_q \sim 300$ MeV. In a
hot medium, this dressing melts and $M_q \to 0$. Since the QCD
Lagrangian for $m_q=0$ is chirally symmetric, $M_q \not= 0$ implies
spontaneous chiral symmetry breaking. The melting $M_q \to 0$ thus
corresponds to chiral symmetry restoration. We shall see later on
that in QCD, as in atomic physics, the shift of the constituent
mass coincides with the onset of conductivity.}
\vspace*{-0.2cm} 
\item{A third type of transition would set in if the attractive
interaction between quarks leads in the deconfined phase to the 
formation of colored bosonic diquark pairs, the Cooper pairs of QCD.
These diquarks can then condense at low temperature to form a color
superconductor. Heating will dissociate the diquark pairs and turn
the color superconductor into a normal color conductor.}
\end{itemize}
Using the baryochemical potential $\mu$ as a measure for the
baryon density of the system, we thus expect the phase diagram 
of QCD to have the schematic form shown in Fig.\ \ref{phase}. 
Given QCD as the fundamental theory of strong interactions, we
can use the QCD Lagrangian as dynamics input to derive the resulting 
thermodynamics of strongly interacting matter. For vanishing 
baryochemical potential, $\mu=0$, this can be evaluated with the
help of the lattice regularisation, leading to finite temperature
lattice QCD. 

\section{From Hadrons to Quarks and Gluons}

Before turning to the results from lattice QCD, we illustrate the 
transition from hadro\-nic matter to quark-gluon plasma by a very simple 
model. For an ideal gas of massless pions, the pressure as function of 
the temperature is given by the Stefan-Boltzmann form
\be
P_{\pi} = 3 {\pi^2 \over 90}~\! T^4 \label{2.3a}
\ee
where the factor 3 accounts for the three charge states of the pion.
The corresponding form for an ideal quark-gluon plasma with two
flavours and three colours is
\be
P_{qg} = \{ 2 \times 8 + {7\over 8}(3 \times 2 \times 2 \times 2) \}
{\pi^2 \over 90}~\! T^4 - B = 
37~\! {\pi^2 \over 90}~\! T^4 - B. \label{2.3b}
\ee
In Eq.\ (\ref{2.3b}), the first temperature term in the curly brackets
accounts for the two spin and eight colour degrees of freedom of the
gluons, the second for the three colour, two flavour, two spin and two
particle-antiparticle degrees of freedom of the quarks, with 7/8 to
obtain the correct statistics. The bag pressure $B$ takes into account
the difference between the physical vacuum and the ground state for
quarks and gluons in a medium.

\medskip

\begin{figure}[htb]
\mbox{
\hskip1.3cm
\epsfig{file=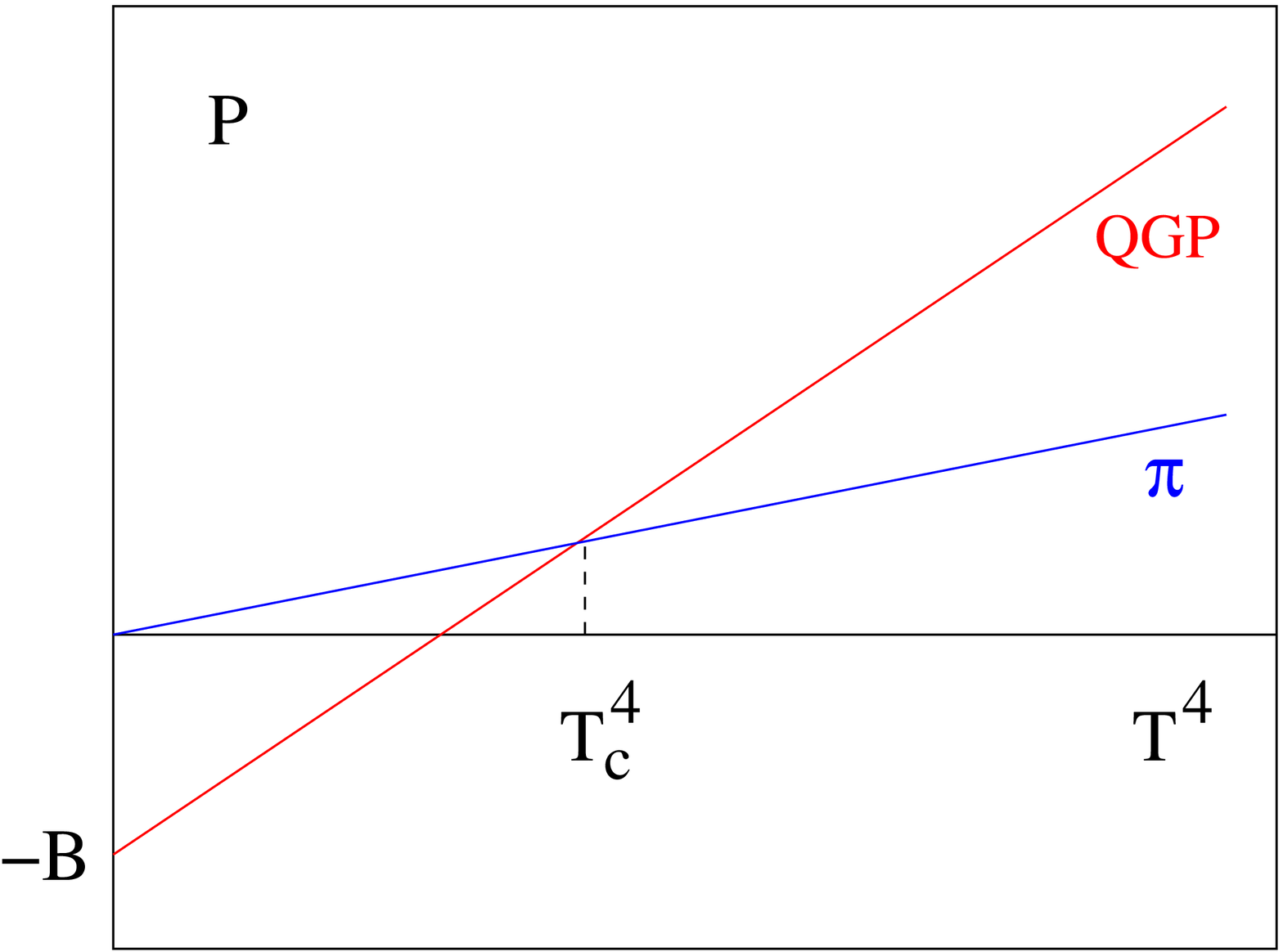,width=5.4cm}
\hskip2.5cm
\epsfig{file=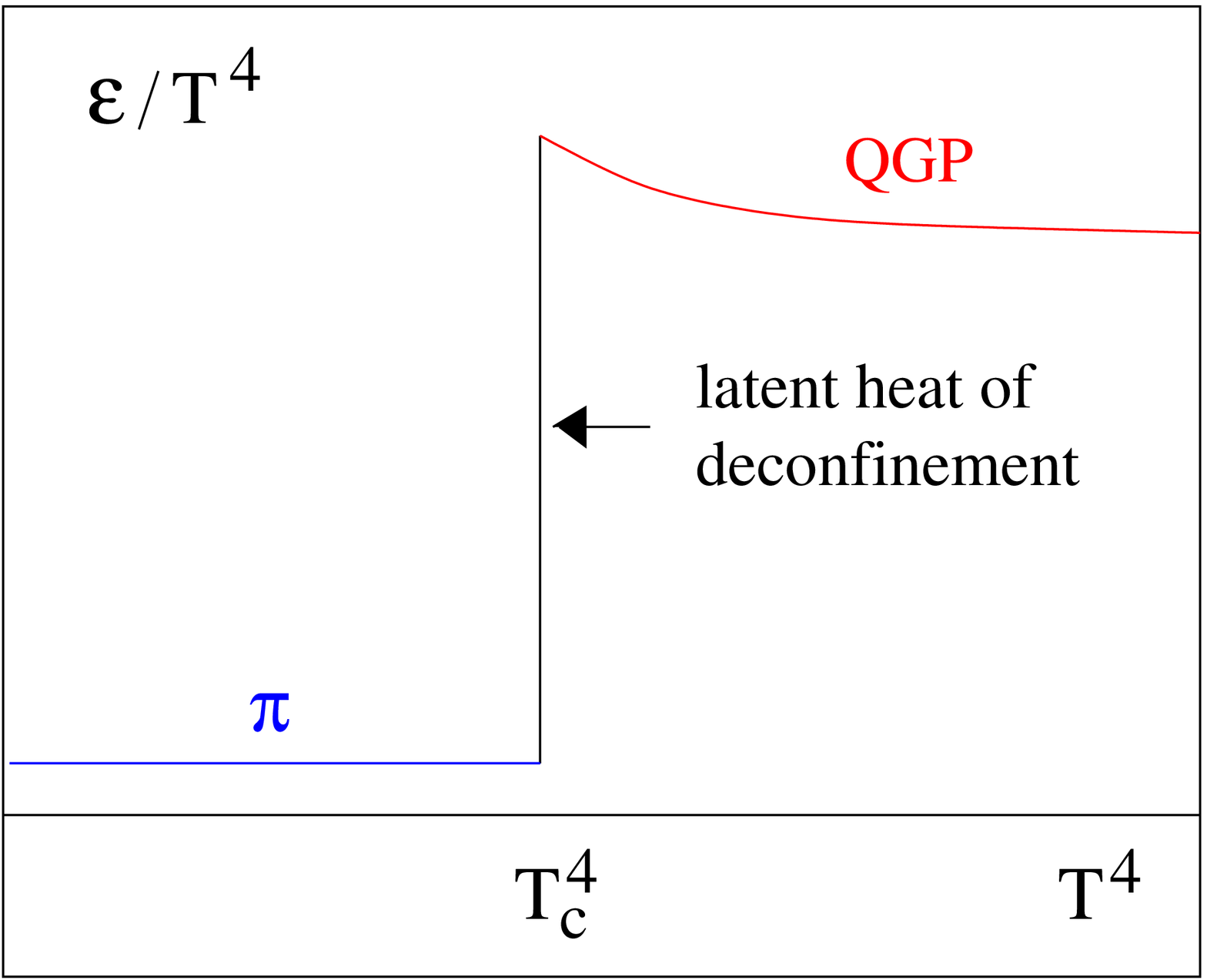,width=5cm}}
\vskip0.5cm
~~~~\hskip3.5cm (a) \hskip7cm (b) 
%\vskip0.1cm
\caption{Pressure and energy density in a two-phase ideal gas
model.}
\label{2_3}
\end{figure}

Since in thermodynamics, a system chooses the state of lowest free
energy and hence highest pressure, we compare in Fig.\ \ref{2_3}$~\!$a the
temperature behaviour of Eq's.\ (\ref{2.3a}) and (\ref{2.3b}). Our
simple model thus leads to a two-phase picture of strongly interacting
matter, with a hadronic phase up to
\be
T_c = \left( {45 \over 17 \pi^2} \right)^{1/4} B^{1/4}
 \simeq 0.72~B^{1/4} \label{2.3e}
\ee
and a quark gluon plasma above this critical temperature. From hadron
spectroscopy, the bag pressure is given by $B^{1/4} \simeq 0.2$ GeV,
so that we obtain
\be
T_c \simeq 150~{\rm MeV} \label{2.3f}
\ee
as the deconfinement temperature. In the next section we shall find
this simple estimate to be remarkably close to the value obtained in
lattice QCD.

\bigskip

The energy densities of the two phases of our model are given by
\be
\e_{\pi} = {\pi^2 \over 10}~\! T^4 \label{2.3g}
\ee
and
\be
\e_{qg} = 37 {\pi^2 \over 30}~\! T^4 + B. \label{2.3h}
\ee
By construction, the transition is first order, and the resulting
temperature dependence is shown in Fig.\ \ref{2_3}$~\!$b. At $T_c$,
the energy density increases abruptly by the latent heat of
deconfinement. We note that even though both phases consist of massless 
non-interacting constituents, the dimensionless ``interaction measure''
\be
{\e - 3P \over T^4}  = {4~\!B\over T^4} \label{2.3i}
\ee
does not vanish in the quark-gluon plasma, due to the (non-perturbative)
difference between physical vacuum and in-medium QCD ground state 
\cite{Asakawa}.

\section{Finite Temperature Lattice QCD}

We now want to show that the conceptual considerations of the last
section indeed follow from strong interaction thermodynamics as based 
on QCD as the input dynamics. QCD is defined by the Lagrangian
\be
{\cal L}~=~-{1\over 4}F^a_{\mu\nu}F^{\mu\nu}_a~-~\sum_f{\bar\psi}^f
_\alpha(i \gamma^{\mu}\partial_{\mu} + m_f
-g \gamma^{\mu}A_{\mu})^{\alpha\beta}\psi^f_\beta
~,\label{2.4}
\ee
with
\be
F^a_{\mu\nu}~=~(\partial_{\mu}A^a_{\nu}-\partial_{\nu}A^a_{\mu}-
gf^a_{bc}A^b_{\mu}A^c_{\nu})~. \label{2.5}
\ee
Here $A^a_{\mu}$ denotes the gluon field of colour $a$ ($a$=1,2,...,8)
and $\psi^f_{\alpha}$ the quark field of colour $\alpha$
($\alpha$=1,2,3) and flavour $f$; the input (`bare') quark masses are
given by $m_f$. With the dynamics thus determined, the corresponding
thermodynamics is obtained from the partition function, which is
most suitably expressed as a functional path integral,
\be
Z(T,V) = \int ~dA~d\psi~d{\bar\psi}~
\exp~\left(-\int_V d^3x \int_0^{1/T} d\tau~
{\cal L}(A,\psi,{\bar\psi})~\right), \label{2.6}
\ee
since this form involves directly the Lagrangian density defining the
theory. The spatial integration in the exponent of Eq.\ (\ref{2.6}) is
performed over the entire spatial volume $V$ of the system; in the
thermodynamic limit it becomes infinite. The time component $x_0$ is
``rotated" to become purely imaginary, $\tau = ix_0$, thus turning the
Minkowski manifold, on which the fields $A$ and $\psi$ are originally
defined, into a Euclidean space. The integration over $\tau$ in Eq.\
(\ref{2.6}) runs over a finite slice whose thickness is determined by
the temperature of the system.

\medskip

From $Z(T,V)$, all thermodynamical observables can be calculated in
the usual fashion. Thus
\be
\epsilon = (T^2/V)\left({\partial \ln Z \over \partial T}\right)_V
\label{2.7}
\ee
gives the energy density, and
\be
P = T \left({\partial \ln Z\over \partial V}\right)_T
\label{2.8}
\ee
the pressure. For the study of critical behaviour, long range
correlations and multi-particle interactions are of crucial importance;
hence perturbation theory cannot be used. The necessary
non-perturbative regularisation scheme is provided by the lattice
formulation of QCD \cite{Wilson}; it leads to a form which can be
evaluated numerically by computer simulation \cite{Creutz}.

\medskip

The calculational methods and techniques of finite temperature lattice QCD 
form a challenging subject on its own, which certainly surpasses the scope 
of this survey. We therefore restrict ourselves here to a summary of the 
main results obtained so far; for more details, we refer to  
excellent recent surveys and reviews \cite{lattice}.

\medskip

The first variable considered in finite temperature lattice QCD 
is the deconfinement measure provided by the Polyakov loop \cite{Larry,Kuti}
\be
L(T) \sim \lim_{r \to \infty}~\exp\{-V(r)/T\} 
\label{polya}
\ee
where $V(r)$ is the potential between a static quark-antiquark pair
separated by a distance $r$. In pure gauge theory, without light quarks,
$V(r) \sim \sigma r$, where $\sigma$ is the string tension; hence here 
$V(\infty)= \infty$ , so that $L=0$. In a deconfined medium, colour
screening among the gluons leads to a melting of the string, which makes 
$V(r)$ finite at large $r$; hence now $L$ does not vanish. It thus becomes
an `order parameter' like the magnetisation in the Ising model: for
the temperature range $0 \leq T \leq T_c$, we have $L=0$ and hence 
confinement, while for $T_c < T$ we have $L>0$ and deconfinement.
The temperature $T_c$ at which $L$ becomes finite thus defines the
onset of deconfinement.

\medskip

In the large quark mass limit, QCD reduces to pure $SU(3)$ gauge theory, 
which is invariant under a global $Z_3$ symmetry. The Polyakov loop provides 
a measure of the state of the system under this symmetry: it vanishes for 
$Z_3$ symmetric states and becomes finite when $Z_3$ is spontaneously 
broken. Hence the critical behaviour of $SU(3)$ gauge theory is in the 
same universality class as that of $Z_3$ spin theory (the 3-state Potts 
model): both are due to the spontaneous symmetry breaking of a global 
$Z_3$ symmetry \cite{Svetitsky}.

\medskip

For finite quark mass $m_q$, $V(r)$ remains finite for $r \to \infty$,
since the `string' between the two colour charges `breaks' when the
corresponding potential energy becomes equal to the mass $M_h$ of the
lowest hadron; beyond this point, it becomes energetically more
favourable to produce an additional hadron. Hence now $L$ no longer
vanishes in the confined phase, but only becomes exponentially small
there,
\be
L(T) \sim \exp\{-M_h/T\}; 
\label{break}
\ee
here $M_h$ is of the order of the $\rho$-mass, so that $L \sim
10^{-2}$, rather than zero. Deconfinement is thus indeed much like the
insulator-conductor transition, for which the order parameter, the
conductivity $\sigma(T)$, also does not really vanish for $T>0$, but
with $\sigma(T) \sim \exp\{-\Delta E/T\}$ is only exponentially small,
since thermal ionisation (with ionisation energy $\Delta E$) produces
a small number of unbound electrons even in the insulator phase.

\medskip

\begin{figure}[htb]
\vspace*{-1cm}
\mbox{\hspace*{1cm}
\epsfig{file=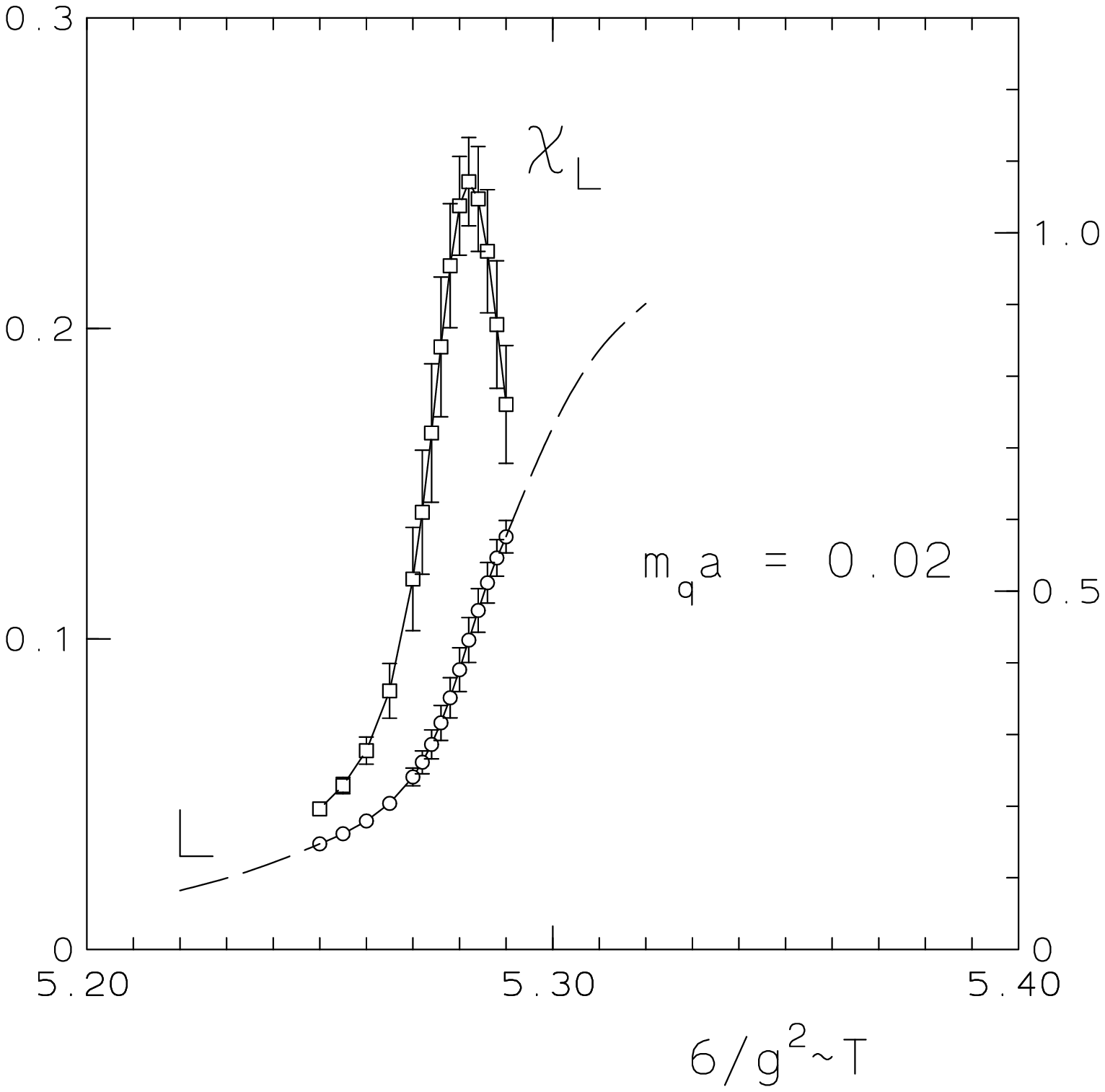,width=6cm}
\hskip1.5cm
\epsfig{file=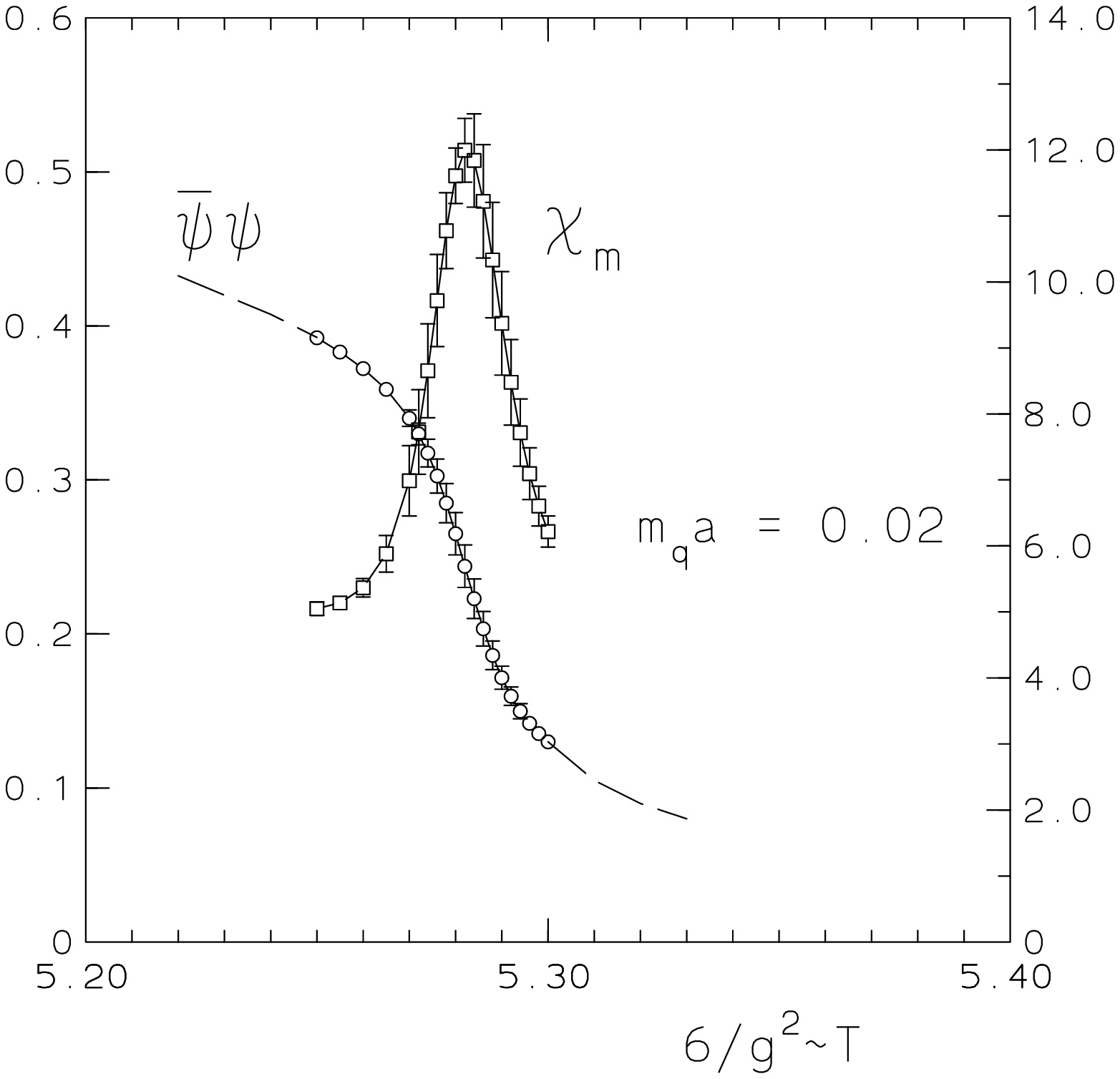,width=6cm}}
\vskip-1.5cm
~~~~\hskip3.5cm (a) \hskip7cm (b) 
\vskip0.1cm
\caption{Polyakov loop and chiral condensate in two-flavour
QCD \cite{K&L}} 
\label{2_4}
\end{figure}

Fig.\ \ref{2_4}$~\!$a shows recent lattice results for $L(T)$ and the
corresponding susceptibility $\x_L(T) \sim \langle L^2 \rangle - 
\langle L \rangle^2$ \cite{K&L}. The calculations were performed 
for the case of two flavours of light quarks, using a current quark 
mass about four times larger than that needed for the physical pion mass.
We note that $L(T)$ undergoes the expected sudden increase from a small
confinement to a much larger deconfinement value. The sharp peak of
$\chi_L(T)$ defines quite well the transition temperature $T_c$, which
we shall shortly specify in physical units.

\vskip-0.5cm

The next quantity to consider is the effective quark mass; it is
measured by the expectation value of the corresponding term in the
Lagrangian, $\langle {\bar \psi} \psi \rangle(T)$. In the
limit of vanishing current quark mass, the Lagrangian becomes chirally
symmetric and $\langle {\bar \psi} \psi \rangle(T)$ the corresponding
order parameter. In the confined phase, with effective constituent quark
masses $M_q \simeq 0.3$ GeV, this chiral symmetry is
spontaneously broken, while in the deconfined phase, at high enough
temperature, we expect its restoration. In the real world, with finite
pion and hence finite current quark mass, this symmetry is also only
approximate, since $\langle {\bar \psi} \psi \rangle (T)$ now never
vanishes at finite $T$.

\medskip

The behaviour of $\langle {\bar \psi} \psi \rangle(T)$ and the
corresponding susceptibility $\chi_m \sim \partial \langle {\bar \psi}
\psi \rangle / \partial m_q$ are shown in Fig.\ \ref{2_4}$~\!$b \cite{K&L},
calculated for the same case as above in Fig.\ \ref{2_4}$~\!$a. 
We note here the
expected sudden drop of the effective quark mass and the associated
sharp peak in the susceptibility. The temperature at which this occurs
coincides with that obtained through the deconfinement measure. We
therefore conclude that at vanishing baryon number density, quark
deconfinement and the shift from constituent to current quark mass
coincide. 

\medskip

We thus obtain for $\mu_B=0$ a rather well defined phase structure,
consisting of a confined phase for $T < T_c$, with $L(T) \simeq 0$ and
$\langle {\bar \psi} \psi \rangle(T) \not= 0$, and a
deconfined phase for $T>T_c$ with $L(T)\not= 0$ and
$\langle {\bar \psi} \psi \rangle(T) \simeq 0$. The
underlying symmetries associated to the critical behaviour at $T=T_c$,
the $Z_3$ symmetry of deconfinement and the chiral symmetry of the quark
mass shift, become exact in the limits $m_q \to \infty$ and $m_q \to
0$, respectively. In the real world, both symmetries are only
approximate; nevertheless, we see from Fig.\ \ref{2_4} 
that both associated measures retain an almost critical behaviour.

\medskip

\begin{figure}[h]
\vskip0.5cm
\begin{minipage}[t]{7.5cm}
\hskip0.3cm 
\epsfig{file=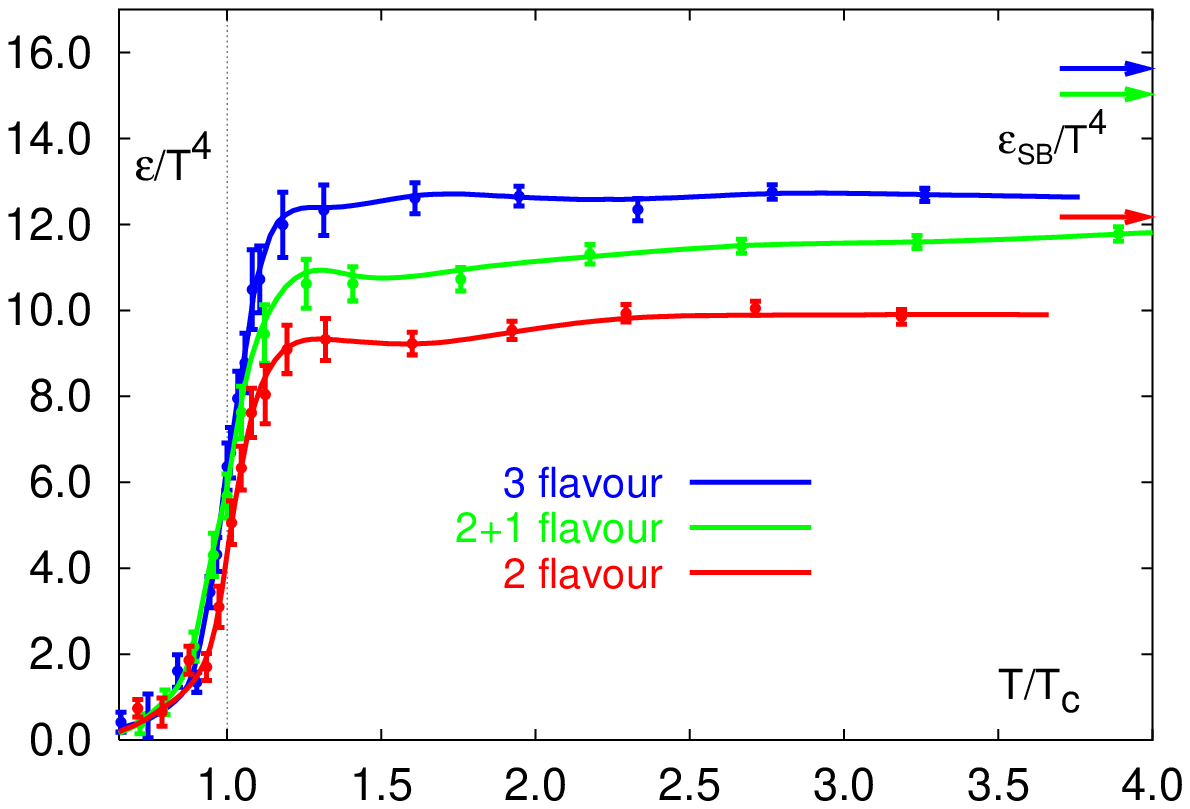,width=6.5cm}
%\vskip-0.1cm
\caption{Energy density vs.\ temperature \cite{Biele}}
\label{edens}
\end{minipage}
\hspace{1cm}
\begin{minipage}[t]{7.5cm}
\vspace*{-4.3cm}
\hspace*{1cm}
\epsfig{file=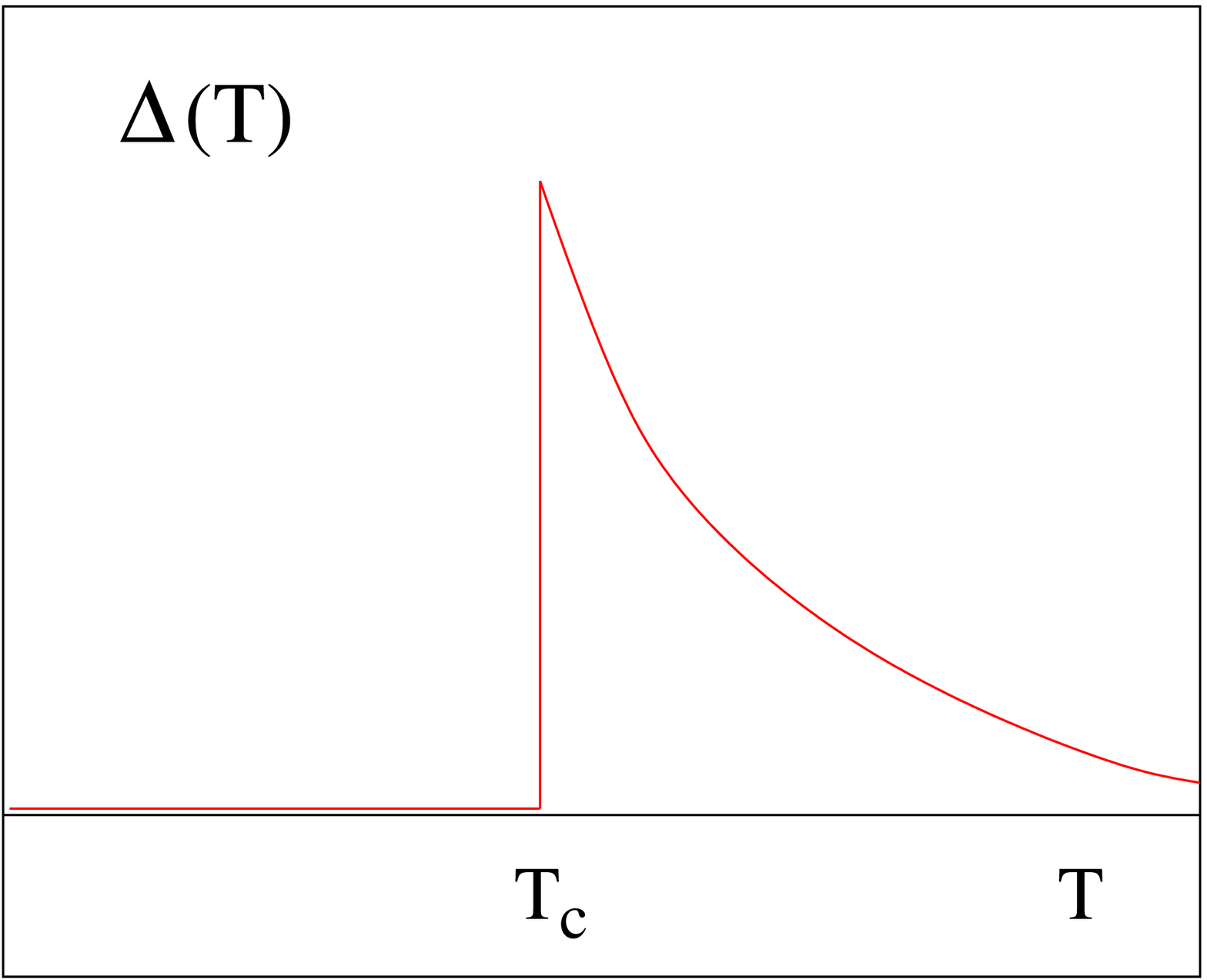,width=5.05cm}
\vskip0.2cm
\caption{Interaction measure vs.\ temperature \cite{Biele}} 
\label{inter}
\end{minipage}
\end{figure}

Next we come to the behaviour of energy density $\e$ and pressure $P$ at
deconfinement \cite{thermo}. In Fig.\ \ref{edens}, it is seen that $\e/T^4$ 
changes quite
abruptly at the above critical temperature $T_c$,
increasing from a low hadronic value to one slightly below that
expected for an ideal gas of massless quarks and gluons \cite{Biele}. 

\vskip-0.5cm

Besides the sudden increase at deconfinement, there are two further points 
to note. In the region $T_c\!<\! T\! <\! 2~T_c$, there still remain strong
interaction effects. As seen in Fig.\ \ref{inter}, the `interaction measure' 
$\Delta=(\e - 3P)/T^4$ remains sizeable and does not vanish, as it would 
for an ideal gas of massless constituents. In the simple model of the 
previous section, such an effect arose due to the bag pressure, and in actual 
QCD, one can also interpret it in such a fashion \cite{Asakawa}. It has
also been considered in terms of a gradual onset of deconfinement starting 
from high momenta \cite{interaction}, and most recently as a possible 
consequence of coloured ``resonance'' states \cite{Shuryak}.
The second point to note is that the thermodynamic observables remain 
about 10 \% below their Stefan-Boltzmann values (marked ``SB'' in Fig.\ 
\ref{edens}) even at very high temperatures, where the interaction measure
becomes very small. Such deviations from ideal gas 
behaviour can be expressed to a large extent in terms of effective `thermal' 
masses $m_{\rm th}$ of quarks and gluons, with $\m \simeq g(T)~T$
\cite{ERSW} - \cite{Patkos}. Maintaining the next-to-leading order term
in mass in the Stefan-Boltzmann form gives for the pressure
\be
P = c~\!T^4 \left[1 - a\left({\m\over T}\right)^2\right] = 
c~T^4 [1 - a~g^2(T)]
\ee
and for the energy density 
\be
\e = 3~\!c~\!T^4 \left[1 - {a\over 3}\left({\m\over T}\right)^2 -
{2a\over 3}\left({\m\over T}\right)\left({d\m\over dT}\right)\right]
=3~\!c~\!T^4 \left[1 - a~\!g^2(T)
- {2 a~\! \m \over 3} \left({dg\over dT}\right)\right],
\ee
where $c$ and $a$ are colour- and flavour-dependent positive constants.
Since $g(T) \sim 1/\log T$,
the deviations of $P$ and $\e$ from the massless Stefan-Boltzmann form  
vanish as $(\log~T)^{-2}$, while the interaction measure
\be
\Delta \sim {1 \over (\log~T)^3}
\ee
decreases more rapidly by one power of $\log~T$.

\medskip

Finally we turn to the value of the transition temperature. Since QCD
(in the limit of massless quarks) does not contain any dimensional
parameters, $T_c$ can only be obtained in physical units by expressing
it in terms of some other known observable which can also be calculated
on the lattice, such as the $\rho$-mass, the proton mass, or the string
tension. In the continuum limit, all different ways should lead to the
same result. Within the present accuracy, they define the uncertainty
so far still inherent in the lattice evaluation of QCD. Using the
$\rho$-mass to fix the scale leads to $T_c\simeq 0.15$ GeV, while
the string tension still allows values as large as $T_c \simeq 0.20$
GeV. Very recently, fine structure charmonium calculations 
(the mass splitting between \J, \X~ and \P) have been used to fix 
the dimensional scale, leading to \cite{BBC} to $T_c \simeq 190
\pm 10$ MeV. In any case, energy densities of some 1 - 2 GeV/fm$^3$ are
needed in order to produce a medium of deconfined quarks and gluons.

\medskip

In summary, finite temperature lattice QCD at vanishing baryon density shows
\begin{itemize}
\vspace*{-0.2cm}
\item{that there is a transition leading to colour deconfinement
coincident with chiral symmetry restoration at $T_c \simeq$ 0.15 - 0.20 GeV;}
\vspace*{-0.2cm}
\item{that this transition is accompanied by a sudden increase in
the energy density  (the ``latent heat of deconfinement") from a small
hadronic value to a much larger value, about 10 \% below
that of an ideal quark-gluon plasma.}
\vspace*{-0.2cm}
\end{itemize}
In the following section, we want to address in more detail the
nature of the critical behaviour encountered at the transition.

\section{The Nature of the Transition}

\begin{figure}[h]
\hspace*{0.5cm}
\begin{minipage}[t]{6cm}
%\hskip0.5cm
\epsfig{file=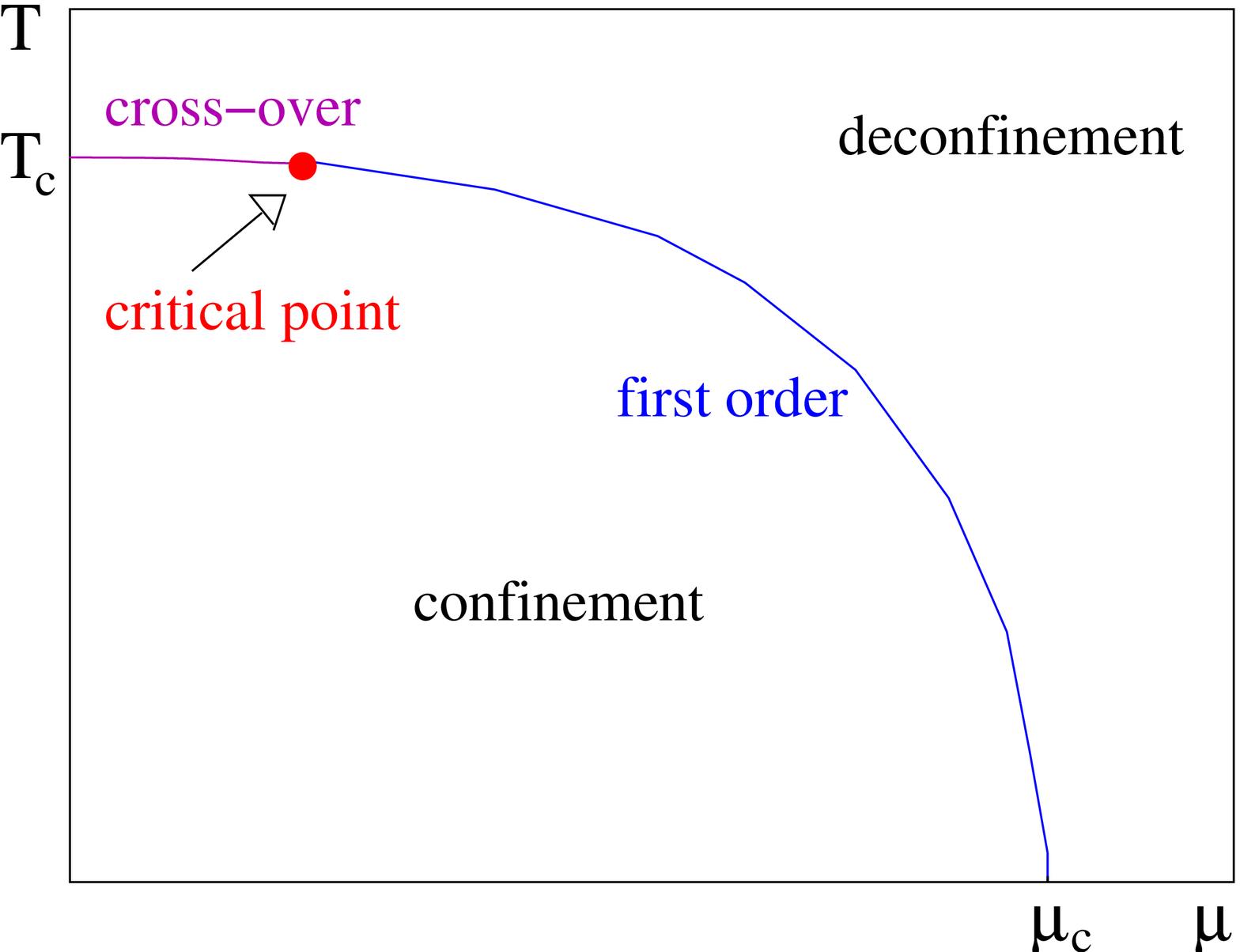,width=6.0cm,height=4.7cm}
\caption{Phase structure in terms of the baryon density}
\label{phase-d}
\end{minipage}
\hspace{2.2cm}
\begin{minipage}[t]{6cm}
\vspace{-4.8cm}
\epsfig{file=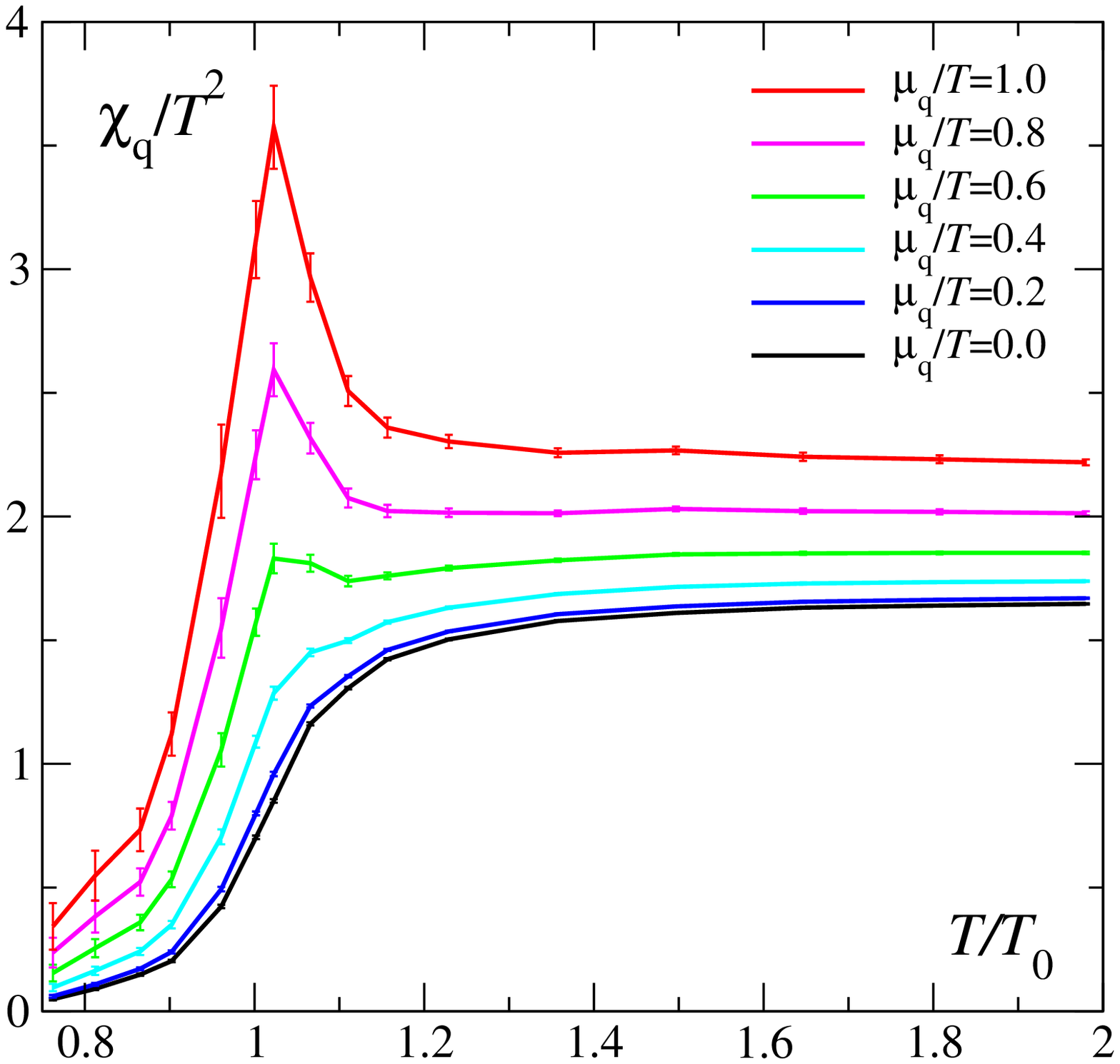,width=6.0cm,height=4.6cm}
\vskip0.2cm
\caption{Baryon number susceptibility $\chi_q$ vs.\ temperature} 
\label{fluc}
\end{minipage}
\end{figure}

We begin with the behaviour for vanishing baryon density ($\mu=0$) and 
come to $\mu\not=0$ at the end. Consider the case of three quark species, 
$u,~d,~s$.
\begin{itemize}
%\vspace*{-0.2cm}
\item{In the limit $m_q \to \infty$ for all quark species, we recover 
pure $SU(3)$ gauge theory, with a deconfinement phase transition provided 
by spontaneous $Z_3$ breaking. It is first order, as is the case for the 
corresponding spin system, the 3-state Potts model.}
\vspace*{-0.5cm}
\item{For $m_q \to 0$ for all quark masses, the Lagrangian becomes 
chirally symmetric, so that we have a phase transition 
corresponding to chiral symmetry restoration. In the case of three
massless quarks, the transition is also of first order.}

\vspace*{0.2cm}

\begin{figure}[htb]
\centerline{\psfig{file=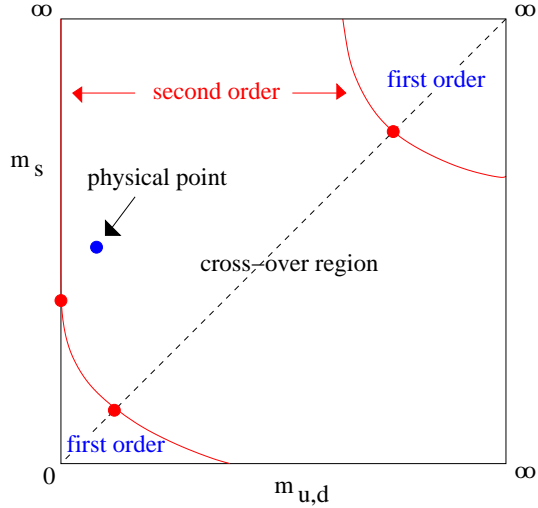,width=7cm}}
\caption{The nature of thermal critical behaviour in QCD}
\label{nature}
\end{figure}
%\vspace*{-0.2cm}

\item{For $0 < m_q < \infty$, there is neither spontaneous $Z_3$
breaking nor chiral symmetry restoration. Hence in general, there is
no singular behaviour, apart from the transient disappearence of the
first order discontinuities on a line of second order transitions.
Beyond this, there is no genuine phase transition, but
only a ``rapid cross-over'' from confinement
to deconfinement. The overall behaviour is summarized in Fig.\ \ref{nature}.} 
\vspace*{-0.2cm}
\item{As already implicitely noted above, both ``order parameters''
$L(T)$ and $\chi(T)$ nevertheless
show a sharp temperature variation for all values
of $m_q$, so that it is in fact possible to define quite well a 
common cross-over point $T_c$.}
\vspace*{-0.2cm}
\item{The nature of the transition thus depends quite sensitively
on the number of flavours $N_f$ and the quark mass values: it can
be a genuine phase transition (first order or continuous), or just 
a rapid cross-over. The case realized in nature, the ``physical point'', 
corresponds to small $u,~d$ masses and a larger $s$-quark mass. 
It is fairly certain today that this point falls into the cross-over region.}
\vspace*{-0.2cm}
\item{Finally we consider briefly the case of finite baryon density,
$\mu\not=0$., so that the number of baryons exceeds that of antibaryons.
Here the conventional computer algorithms of lattice QCD break down,
and hence new calculation methods have to be developed. First such attempts
(reweighting \cite{Fodor}, analytic continuation \cite{Lombardo}, 
power series \cite{Swansea}) suggest for two light 
quark flavours the phase diagram shown in Fig.\ \ref{phase-d}.
It shows non-singular 
in a region between $0 \leq \mu < \mu_t$, a tricritical point at $\mu_t$, 
and beyond this a first order transition. 
Recent lattice calculations provide some support for such behaviour;
as shown in Fig.\ \ref{fluc}, the baryon density fluctuations appear to
diverge for some critical value of the baryochemical potential 
\cite{Swansea}.}

\end{itemize}

\medskip

The critical behaviour for strongly interacting matter at low or
vanishing baryon density, describing the onset of confinement in 
the early universe and in high energy nuclear collisions, thus occurs 
in the rather enigmatic form of a ``rapid cross-over''. There is no 
thermal singularity and hence, in a strict sense, there are neither 
distinct states of matter nor phase transitions between them. So what 
does the often mentioned experimental search for a ``new state of matter'' 
really mean? How can a new state appear without a phase transition?
Is there a more general way to define and distinguish different states of bulk
media? After all, in statistical QCD one does find that thermodynamic 
observables -- energy and entropy densities, pressure, as well as 
the ``order parameters'' $L(T)$ and $\chi(T)$ -- continue to change rapidly 
and thus define a rather clear transition line in the entire cross-over 
region. Why is this so, what is the mechanism which causes 
such a transition?
 
\medskip

In closing this section, we consider a speculative answer to this
rather fundamental question \cite{hs-perco}. The traditional phase 
transitions, such as the freezing of water or the magnetization of iron,
are due to symmetry breaking and the resulting singularities of the
partition function. But there are other ``transitions'', such as
making pudding or boiling an egg, where one also has two clearly
different states, but no singularities in the partition function.
Such ``liquid-gel'' transitions are generally treated in terms of
cluster formation and percolation \cite{Stauffer}. They also correspond 
to critical
behaviour, but the quantities that diverge are geometric (cluster
size) and cannot be obtained from the partition function. 

\begin{figure}[h]
\centerline{\psfig{file=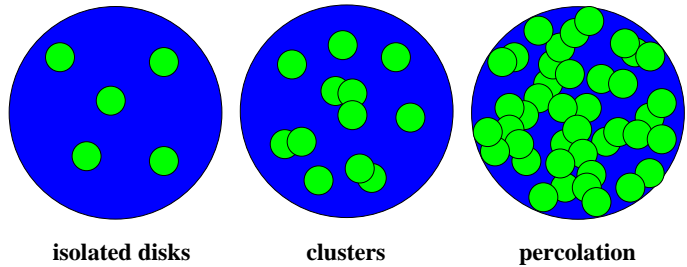,width=9cm}}
\caption{Lilies on a pond}
\label{lilies}
\end{figure}

\medskip

\begin{figure}[htb]
\centerline{\psfig{file=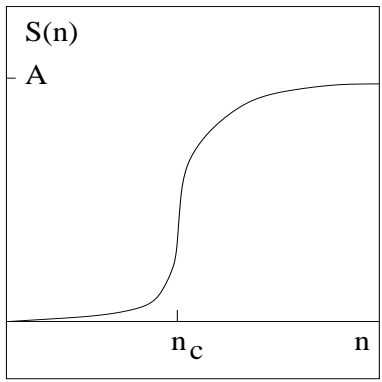,width=5cm,height=5cm}}
\caption{Cluster size $S(n)$ vs.\ density $n$}
\label{cluster}
\end{figure}

The simplest example of this phenomenon is provided by two-dimensional 
disk percolation, something poetically called ``lilies on a pond'' 
(see Fig.\ \ref{lilies}). More formally: one distributes small disks of
area $a=\pi r^2$  randomly on a large surface $A=\pi R^2$, $R\gg r$, with
overlap allowed. With an increasing number of disks, clusters begin to
form. If the large surface were water and the small disks floating water 
lilies: how many lilies are needed for a cluster to connect the opposite 
sides, so that an ant could walk across the pond without getting its feet
wet? Given $N$ disks, the disk density is
$n=N/A$. Clearly, the average cluster $S(n)$ size will increase with $n$.
The striking feature is that it does so in a very sudden way (see
Fig.\ \ref{cluster}); as $n$ approaches some ``critical value'' $n_c$,
$S(n)$ suddenly becomes large enough to span the pond. In fact, in the 
limit $N \to \infty$ and $A \to \infty$ at constant $n$, both
$S(n)$ and $dS(n)/dn$ diverge for $n \to n_c$: we have percolation
as a geometric form of critical behaviour.

\medskip

The critical density for the onset of percolation has been determined
(numerically) for a variety of different systems. In two dimensions, disks 
percolate at $n_c\simeq 1.13/(\pi r^2)$, i.e., when we have a little more 
than one disk per unit area. Because of overlap, at this point only {68\%}
of space is covered by disks, 32\% remain empty. Nevertheless, when
our ant can walk across, a ship can no longer cross the pond, and vice 
versa. This is a special feature of two dimensions (the ``fence effect''),
and no longer holds for $d>2$. 

\medskip

In three dimensions, the corresponding problem is one of overlapping
spheres in a large volume. Here the critical density for the percolating 
spheres becomes $n_c \simeq 0.34/[(4\pi/3)r^3]$, with $r$ denoting the 
radius of the little spheres now taking the place of the small disks we 
had in two dimensions. At the critical point in three dimensions, however, 
only {29\%} of space is covered by overlapping spheres, while 71\% remains
empty, and here both spheres and empty space form infinite connected
networks. If we continue to increase the density of spheres, we reach
a second critical point at $\bar n_c \simeq 1.24/[(4\pi/3)r^3]$, at which
the vacuum stops to form an infinite network: now 71\% of space is covered 
by spheres, and for $n>\bar n_c$, only isolated vacuum bubbles remain.

\medskip

\begin{figure}[htb]
\vspace*{0.2cm}
\centerline{\epsfig{file=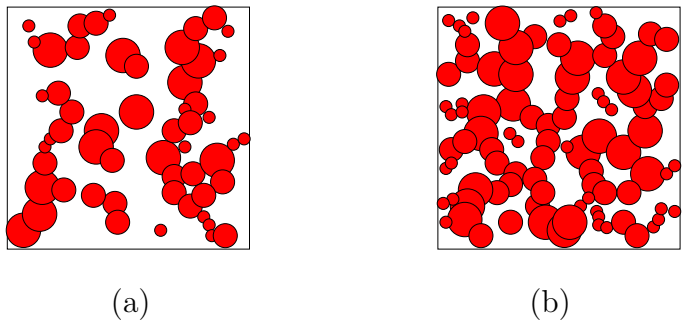,width=9cm}}
\vskip0.3cm
\hskip 4.9cm (a) \hskip 4.9cm (b)
\caption{Hadron and vacuum percolation thresholds}
\label{percotrans}
\end{figure}

Let us then consider hadrons of intrinsic size $V_h=(4\pi/3)r_h^3$, with 
$r_h \simeq 0.8$ fm. In three-dimensional space, the formation of a 
connected large-scale cluster first occurs at the density
\be
n_c= {0.34 \over V_h} \simeq 0.16~{\rm fm}^{-3}.
\label{hadronmatter} 
\ee
This point specifies the onset of hadronic {\sl matter}, in contrast to
a {\sl gas} of hadrons, and it indeed correctly reproduces the density of 
normal nuclear matter. However, at this density the vacuum as connected 
medium also still exists (see Fig.\ \ref{percotrans}a).

\medskip

To prevent infinite connecting vacuum clusters, a much higher hadron 
density is needed, as we saw above. Measured in hadronic size units, 
the vacuum disappears for
\be
\bar{n}_c= {1.24 \over V_h} \simeq 0.56~{\rm fm}^{-3}, 
\label{vacuummatter} 
\ee
schematically illustrated in Fig.\ \ref{percotrans}b. If we assume that at
this point, the medium is of an ideal gas of all known hadrons and hadronic 
resonances, then we can calculate the temperature of the gas at the 
density $\bar{n}_c$: $n_{\rm res}(T_c) = \bar{n}_c$ implies $T_c \simeq 
170$ MeV, which agrees quite well with the value of the deconfinement 
temperature found in lattice QCD for $\mu=0$.
 
\medskip

We can thus use percolation to define the states of hadronic matter.
At low density, we have a hadron gas, which at the percolation point
$n_c$ turns into connected hadronic matter. When this becomes so dense
that only isolated vacuum bubbles survive, at $\bar{n}_c$, it turns into 
a quark-gluon plasma. This approach provides the correct values both for 
the density of standard nuclear matter and for the deconfinement transition 
temperature.

\medskip

Such considerations may in fact well be of a more general nature than
the problem of states and transitions in strong interaction physics. 
The question of whether symmetry or connectivity (cluster formation)
determines the different states of many-body systems has intrigued
theorists in statistical physics for a long time \cite{F-K}. The
lesson learned from spin systems appears to be that cluster formation
and the associated critical behaviour are the more general features, which 
under certain conditions can also lead to thermal criticality, i.e., 
singular behaviour of the partition function. 

\section{Probing the Quark-Gluon Plasma}

We thus find that at sufficiently high temperatures and/or densities,
strongly interacting matter will be in a new state, consisting of deconfined
quarks and gluons. How can we probe the properties of this state, how
can we study its features as function of temperature and density? We want
to address this question here in the sense of Einstein, who told us to
make things as simple as possible, but not simpler. So let us start with
a theorist's experimental set-up, consisting of a volume of unknown
strongly interacting matter and a Bunsen burner, to heat it up and thus
increase its energy density.

\bigskip

\begin{figure}[htb]
\centerline{\psfig{file=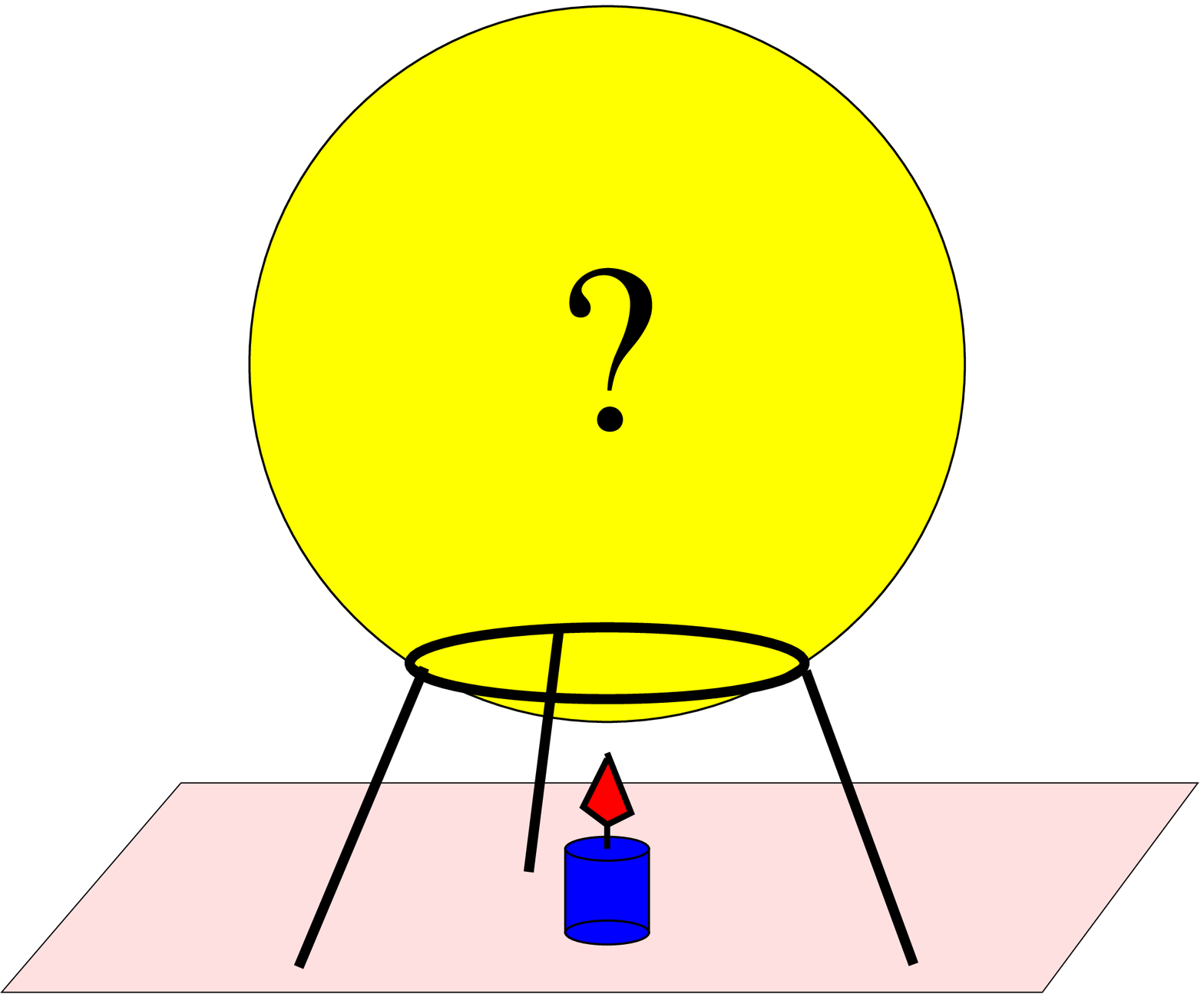,width=5cm}}
\end{figure}

\newpage

There are a number of methods we can use to study the unknown matter
in our container:
\begin{itemize}
\item{hadron radiation,}
\item{electromagnetic radiation,}
\item{dissociation of a passing quarkonium beam,}
\item{energy loss of a passing hard jet.}
\end{itemize}
All methods will be dealt with in detail during the course of this
school. Here we just want to have a brief first look.

\medskip

\begin{figure}[htb]
\centerline{\psfig{file=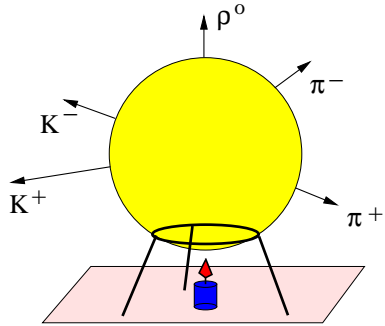,width=5cm}}
\caption{Hadron radiation}
\end{figure}

\medskip

First of all, we note that the unkown medium radiates, since its
temperature is (by assumption) much higher than that of its environement.
Hadron radiation means that we study the emission of
hadrons consisting of light ($u,d,s$) quarks; their size is given by 
the typical hadronic scale of about 1 fm $\simeq$ 1/(200 MeV).
Since they cannot exist inside a deconfined medium, they are formed
at the transition surface between the QGP and the physical vacuum.
The physics of this surface is independent of the interior - the
transition from deconfinement to confinement occurs at a temperature
$T \simeq\ 160- 180$ MeV,
no matter how hot the QGP initially was or still is in the interior of
our volume. This is similar to having hot water vapor inside a glass
container kept in a cool outside environement: at the surface, the
vapor will condense into liquid, at a temperature of 100$^{\circ}$C - no
matter how hot it is in the middle. As a result, studying
soft hadron production in high energy collisions will provide us 
with information about the hadronization transition, but not about
the hot QGP. The striking observation that the relative hadron 
abundances in all high energy collisions are the same, from $e^+e^-$ 
annihilation to hadron-hadron and heavy ion interactions, and that 
they correspond to those of an ideal resonance gas at $T\simeq 170$
MeV, is a direct consequence of this fact \cite{Hagedorn,thermal} . 

\medskip

\begin{figure}[htb]
\centerline{\psfig{file=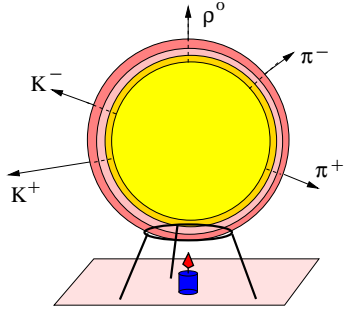,width=4.5cm}}
\caption{Radial flow and hadron radiation}
\label{flow}
\end{figure}

Hadron radiation, as we have pictured it here, is oversimplified from
the point of view of heavy ion interactions. In
the case of static thermal radiation, at the point of hadronization
all information about the earlier stages of the medium is lost, as we
had noted above. If, however, the early medium has a very high energy
density and can expand freely, i.e., is not constrained by the walls
of a container, then this expansion will lead to a global hydrodynamic 
flow \cite{Landau},
giving an additional overall boost in momentum to the produced hadrons:
they will experience a ``radial flow'' depending on the initial energy 
density (see Fig.\ \ref{flow}). Moreover, if the initial conditions were 
not spherically symmetric, as is in fact the cases in peripheral heavy
ion collisions, the difference in pressure in different spatial
directions will lead to a further ``directed'' or ``elliptic'' flow.
Since both forms of flow thus do depend on the initial conditions,
flow studies of hadron spectra can, at least in principle, provide 
information about the earlier, pre-hadronic stages. 

\medskip

\medskip

The unknown hot medium also radiates electromagnetically, i.e., it
emits photons and dileptons ($e^+e^-$ or $\mu^+\mu^-$ pairs)
\cite{e-m}. These
are formed either by the interaction of quarks and/or gluons, or by
quark-antiquark annihilation. Since the photons and leptons interact
only electromagnetically, they will, once they are formed, leave the medium 
without any further modification. Hence their spectra provide information
about the state of the medium at the place or the time they were formed, 
and this can be in its deep interior or at very early stages of its
evolution. Photons and dileptons thus do provide a possible probe of the 
hot QGP. The only problem is that they can be formed anywhere and at any
time, even at the cool surface or by the emitted hadrons. The task in
making them a viable tool is therefore the identification of the hot
``thermal'' radiation indeed emitted by the QGP. 

\medskip

\begin{figure}[htb]
\centerline{\psfig{file=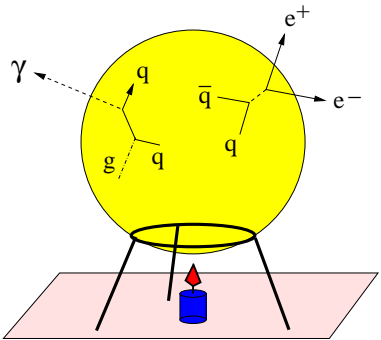,width=5cm}}
\caption{Electromagnetic radiation}
\end{figure}

Both electromagnetic and hadronic radiation are emitted by the medium
itself, and they provide some information about the state of the medium at
the time of emission. Another possible approach is to test the medium
with an outside probe, and here we have two so far quite successful
examples, quarkonia and jets.

\medskip

Quarkonia are a special kind of hadrons, bound states of a heavy ($c$ or
$b$) quark and its antiquark. For the ground states \J~ and \U~the
binding energies are around 0.6 and 1.2 GeV, respectively, and thus
much larger than the typical hadronic scale $\Lambda \sim 0.2$ GeV;
as a consequence, they are also much smaller, with radii of about 0.1 and
0.2 fm. It is therefore expected that they can survive in a quark-gluon
plasma through some range of temperatures above $T_c$, and this is
in fact confirmed by lattice studies \cite{Psi-Lat}. 

\medskip

The higher excited quarkonium
states are less tightly bound and hence larger, although their
binding energies are in general still larger, their radii still smaller, 
than those
of the usual light quark hadrons. Take the charmonium spectrum as
example: the radius of the \J(1S) is about 0.2 fm, that of the
\X(1P) about 0.3 fm, and that of the \P(2S) 0.4 fm. Since deconfinement
is related to colour screening, the crucial quantity for dissociation of
a bound state is the relation of binding to screening radius. We therefore
expect that the different charmonium states have different ``melting
temperatures'' in a quark-gluon plasma. Hence the spectral analysis of
in-medium quarkonium dissociation should provide a QGP thermometer
\cite{MS}. 

\begin{figure}[htb]
\centerline{\psfig{file=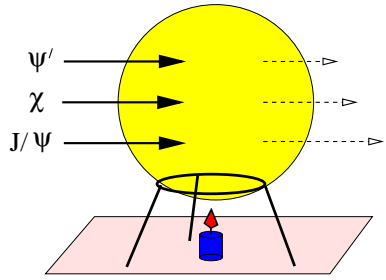,width=5cm}}
\caption{Charmonium suppression}
\label{Jpsi-probe}
\end{figure}

\medskip

\begin{figure}[htb]
\vspace*{0.5cm}
{~~~~~~~~~~~~~~~~~~~~~~\psfig{file=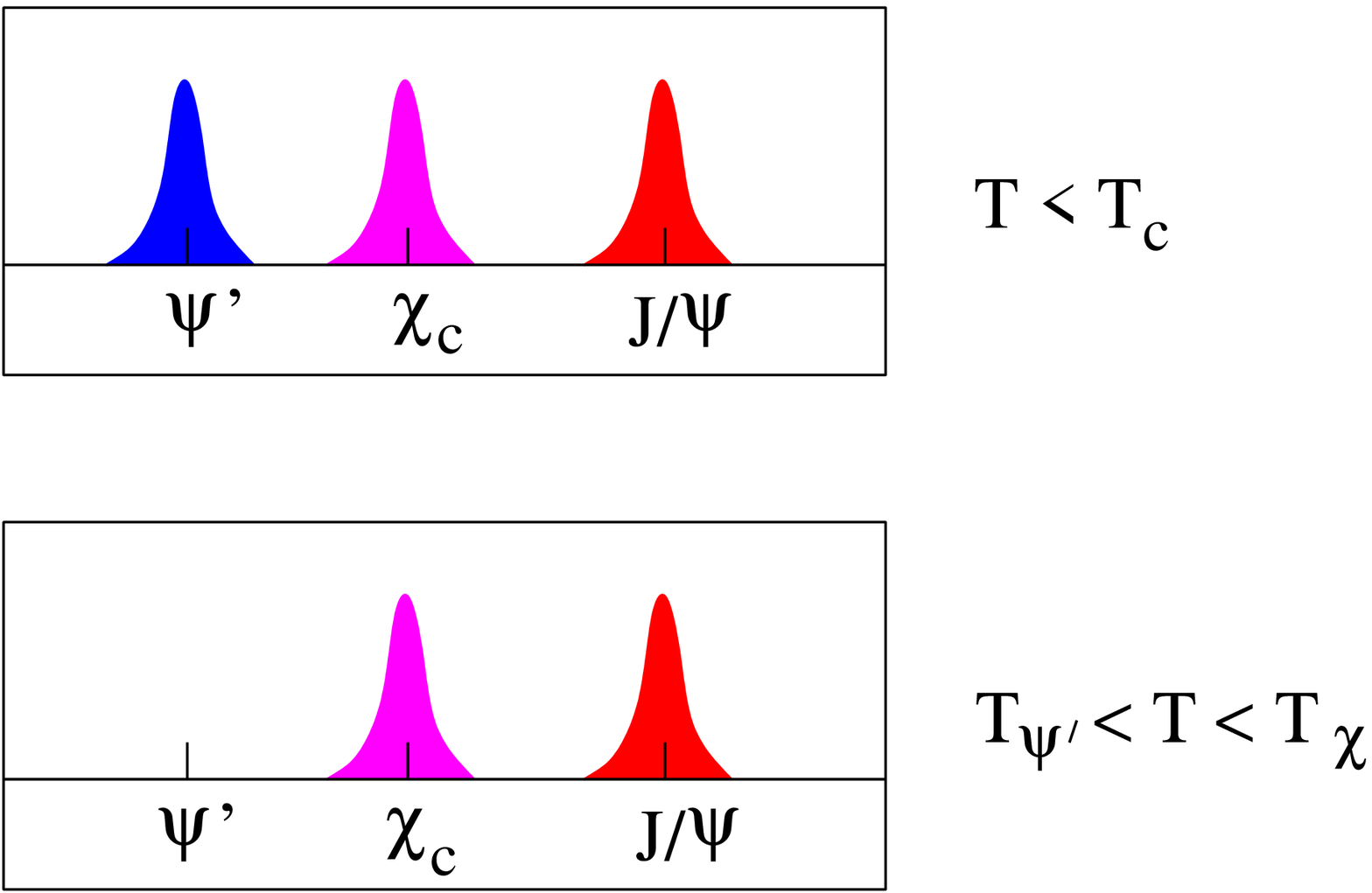,width=5.4cm}}
\hfill 
\vspace*{-3cm}
{\psfig{file=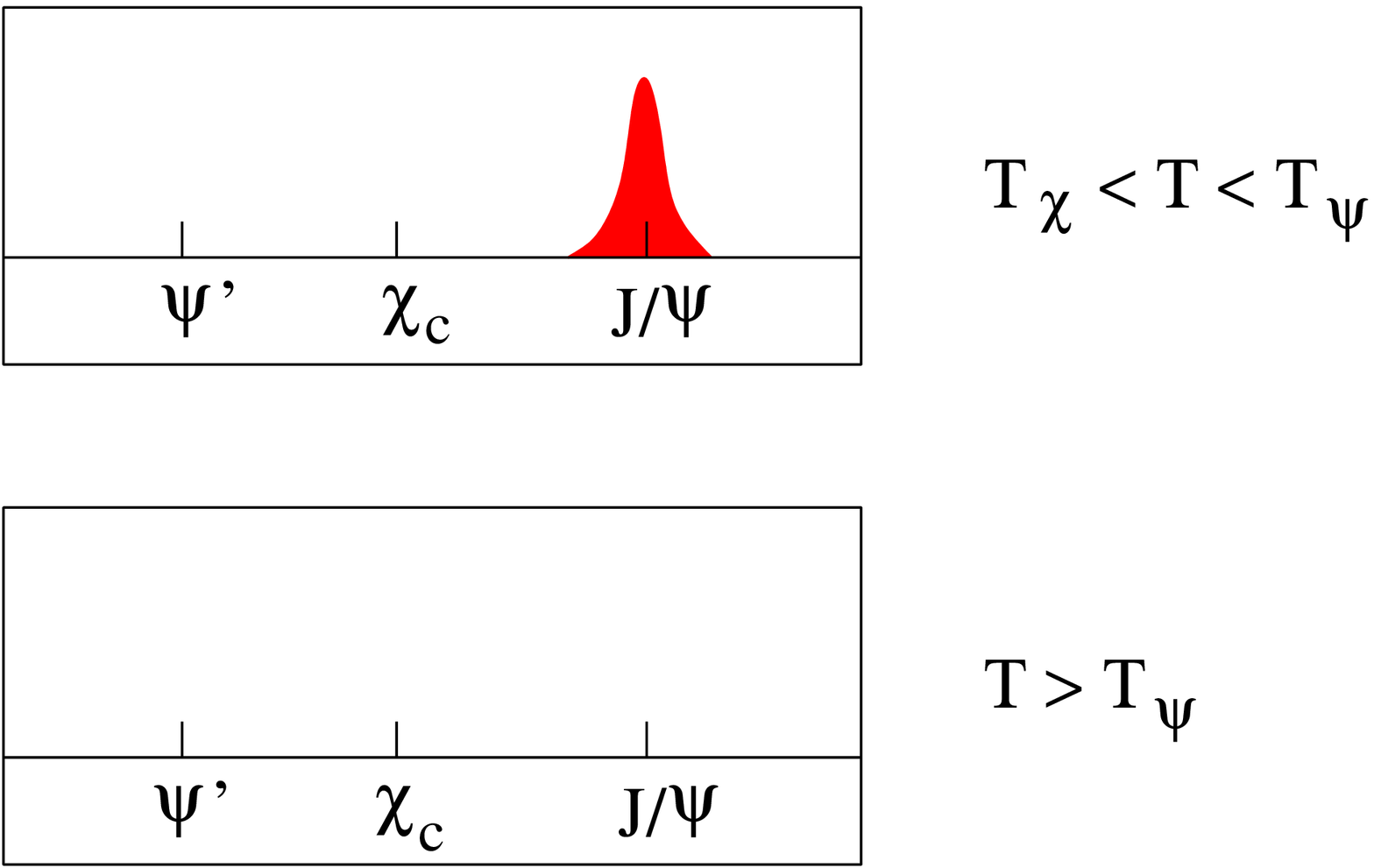,width=5.5cm}~~~~~~~~~}
\vspace*{3.5cm}
\caption{Charmonia as thermometer}
\label{temp}
\end{figure}

As probe, we then shoot beams of specific charmonia (\J,~\X,~\P)
into our medium 
sample and check which comes out on the other side (Fig.\ \ref{Jpsi-probe}).
If all three survive, we have an upper limit on the temperature, and by
checking at just what temperature the \P, the \X~and the \J~are
dissociated, we have a way of specifying the temperature of the 
medium \cite{KS}, as illustrated in Fi.\ \ref{temp}.

\medskip

\medskip

\begin{figure}[htb]
\centerline{\psfig{file=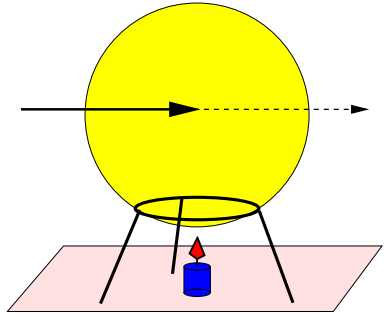,width=5cm}}
\caption{Jet quenching}
\label{jet-probe}
\end{figure}

Another possible probe is to shoot an energetic parton, quark or gluon, 
into our medium to be tested. How much energy it loses when it comes out
on the other side will tell us something about the density of the
medium \cite{Bj}. In particular, the density increases by an order of magnitude
or more in the course of the deconfinement transition, and so the
energy loss of a fast passing colour charge is expected to increase
correspondingly (Fig.\ \ref{jet-probe}). Moreover, for quarks, the amount 
of jet quenching is predicted to depend on the mass of the quarks.  

\medskip

In using quarkonia and jets as tools, we have so far considered a 
simplified situtation, in which we test a given medium with distinct
external probes. In heavy ion collisions, we have to create the probe
in the same collision in which we create the medium.
Quarkonia and jets (as well as open charm/beauty and very energetic dileptons 
and photons) constitute so-called ``hard probes'', whose production 
occurs at the very early stages of the collision, before the medium is
formed; they are therefore indeed present when it appears. Moreover,
their production involves large energy/momentum scales and can 
be calculated by perturbative QCD techniques and tested in $pp/pA$
collisions, so that behaviour and strength of such outside ``beams'' 
or ``colour charges'' are indeed quite well known.

\section{Summary}

We have shown that strong interaction thermodynamics leads to a well-defined
transition from hadronic matter to a plasma of deconfined quarks and gluons.
For vanishing baryon number density, the transition leads to simultaneous
deconfinement and chiral symmetry restoration at $T_c \simeq 160 - 190$ MeV.
At this point, the energy density increases by an order of magnitude through
the latent heat of deconfinement.

\medskip

The properties of the new medium above $T_c$, the quark-gluon plasma, can 
be studied through hard probes (quarkonium and open charm/beauty production, 
jet quenching) and electromagnetic radiation (photons and dileptons). 
Information about transition aspects is provided by light hadron
radiation; through hydrodynamic flow, this can also shed light on 
pre-hadronic features.

\end{document}